# Do Short GRBs Exhibit an Anticorrelation between Their Intrinsic Duration and Redshift?

Ali M. Hasan, Walid J. Azzam

Department of Physics, College of Science, University of Bahrain, Sakhir, Bahrain
Email: wjazzam@uob.edu.bh, wjazzam@gmail.com

## Abstract

Gamma-ray bursts (GRBs) are violent stellar explosions that are traditionally divided into two groups: short bursts (SGRBs) with an observed duration $T_{90} < 2$ s, and long bursts (LGRBs) with an observed duration $T_{90} > 2$ s, where $T_{90}$ refers to the time needed for 90% of the fluence to be detected. Studies of progenitor models suggest that LGRBs emanate from the core collapse of massive stars, while SGRBs result from the merging of two compact objects, like two neutron stars or a neutron star and a black hole. Recent studies have found evidence that there is an anticorrelation between the intrinsic duration and the redshift of long GRBs. In this study, we first check whether LGRBs exhibit an anticorrelation between their intrinsic duration and redshift using an expanded dataset of long bursts that we have compiled. Next, we investigate whether this anticorrelation applies to SGRBs as well using a sample of short GRBs that we have compiled. Our analysis confirms the results obtained by previous studies regarding the anticorrelation for LGRBs. On the other hand, our results indicate that short GRBs do not exhibit such an anticorrelation. We discuss the implications of our results in the context of how metallicity evolves with redshift and the role that it might play in the aforementioned anticorrelation.

## 1. Introduction

The characterization and the classification of gamma-ray bursts (GRBs) have proved a challenge to researchers since GRBs were first discovered in 1967 [1]. Traditionally, bursts have been classified into long GRBs (LGRBs) with $T_{90} > 2$ s and short GRBs (SGRBs) with $T_{90} < 2$ s, where $T_{90}$ is the time that the detectors take to observe 90% of the GRB's fluence [2]. Generally, LGRBs are believed to emanate from the core collapse of massive stars, while SGRBs are thought to be the result of the merger of two compact objects, like two neutron stars (NSs). However, it is important to keep in mind that this association between GRBs and the presumed mechanism behind their formation is currently being reexamined [3]-[7]. One reason for this recent debate regarding the sources of long and short GRBs is that $T_{90}$ represents the observed duration, not the intrinsic one, thus it may be influenced by cosmological effects, meaning that it does not directly and necessarily convey the intrinsic physical properties of GRBs. Therefore, many studies have pushed for newer classifications that better represent the nature of GRBs [8]-[12].

Despite this debate, the traditional classification of bursts into LGRBs and SGRBs does have some merit because each class does exhibit some distinguishing features and characteristics. For example, several studies have shown that the relation between equivalent isotropic energy, $E_{iso}$, and the intrinsic peak energy, $E_{p,i}$, which is known as the Amati relation, exhibits different correlation strengths and coefficients for SGRBs and LGRBs, to the degree that some studies have suggested employing the Amati relation as a discriminator between the two classes of gamma-ray bursts [13]-[19].

Another relation that has recently attracted some attention in the literature is the anticorrelation between the intrinsic duration, $T_{int}$, of GRBs and their redshift $z$, where $T_{int}$ is obtained from the observed duration as follows: $T_{int} = T_{90}/(1 + z)$. A series of studies have found significant evidence that there is an anticorrelation between $T_{int}$ and $z$ for LGRBs [20]-[23]. The aim of this brief paper is twofold: first, to check and confirm the results of these previous studies but for an expanded data sample of long bursts; second, to investigate whether short GRBs exhibit an anticorrelation between $T_{int}$ and $z$, like the one seen for LGRBs. In Section 2, we describe the method that we employed. In Section 3, we present the data sets that we compiled and used, both for LGBRs and SGRBs, and the results that we obtained, and we also provide a discussion of our results. Our conclusion is provided in Section 4.

## 2. Method

Our first step involved compiling the redshift and duration data for both long and short bursts. For the LGRBs, we compiled our dataset, which is presented in **Table A1** in **Appendix**, from different published sources [24]-[26]. On the other hand, we compiled our SGRB dataset, which is presented in **Table A2** in **Appendix**, from the recent study by Zhu *et al.* [27] in which the authors provided a systematic analysis of the intrinsic properties of short bursts.

After compiling the data, we binned them and carried out a best-fit of the form:

$$\log T_{int} = a \log(1+z) + b. \quad (1)$$

The LGRB dataset, which consists of 173 bursts, was binned into 16 bins (15 bins with 11 LGRBs and 1 bin with 8 LGRBs). In a similar fashion, we binned the SGRB data sample, which consists of 83 bursts, into 12 bins (11 bins with 7 SGRBs and 1 bin with 6 SGRBs).

## 3. Results and Discussion

The best fit that we obtained for the LGRB dataset is shown in **Figure 1**. The fitting gives $T_{int} \propto (1+z)^{-1.24 \pm 0.34}$ ($\chi^2/dof = 1.54$) with a p-value much less than the standard 1% - 5% for rejecting the null hypothesis. Moreover, the fitting value $\alpha = -1.24 \pm 0.34$ is within $1\sigma$ of those that are found in the previous studies, thus giving further support to the anticorrelation claims for LGRBs [20]-[23].

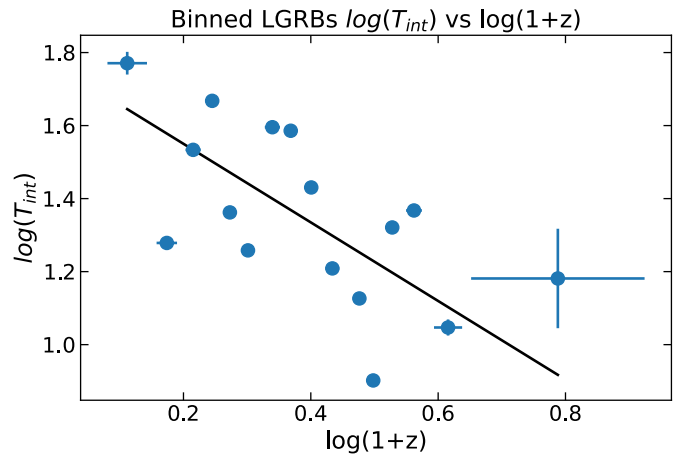

**Figure 1.** Shows the $T_{int}$ vs $(1+z)$ fitting for LGRBs. The best fit line in black is given by $T_{int} \propto (1+z)^{-1.24 \pm 0.34}$ with the reduced $\chi^2$ being 1.54 (p-value = $4 \times 10^{-8}$).

For the SGRB dataset, our fitting is shown in **Figure 2**. The data reveal that there is no clear relationship between the intrinsic duration and the redshift for SGRBs, as reflected by the high reduced $\chi^2$ value and the high scatter in the data points.

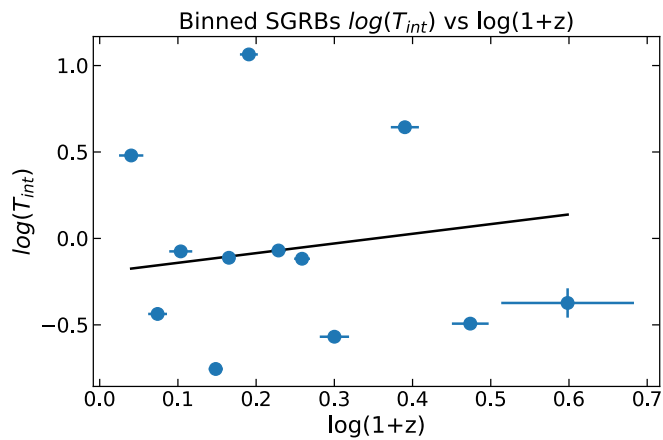

**Figure 2.** Shows the $T_{int}$ vs $(1+z)$ fitting for SGRBs. The best fit line in black is given by $T_{int} \propto (1+z)^{0.56 \pm 1.75}$ with the reduced $\chi^2$ being 25.74 (p-value = 0.01).

One possible explanation of our results is that it is due to the role that metallicity plays. Some studies have suggested that metallicity can be linked to this anticorrelation, as metallicity itself decreases with increasing redshift [24]. The study by [24] presents two scenarios for LGRB progenitors, a single massive star and an interacting binary system involving massive stars. In the case of a single massive star, more metallic stars have larger stellar radii, which produces larger accretion disks, hence longer durations during core-collapse. On the other hand, for interacting binary systems with massive stars, the anticorrelation comes from metallicity affecting the mass-loss, which in turn, influences the binary separation and the angular momentum, and thus the intrinsic duration of the energy release. Admittedly, the link to metallicity is hard to establish and confirm due to the paucity of GRB data points that have observed metallicities. The large uncertainty in metallicity plus the scarcity of metallicity data for GRBs presents a challenge when it comes to incorporating its effect on GRB intrinsic correlations, though this has not inhibited theorists from putting forward plausible models [24]. Proper and extensive identification of GRB host galaxies and the improvement and refinement of metallicity observations is needed to push this field of study further. Such extensive and systematic observations will help clarify the potential role and impact that metallicity has on GRB durations and intrinsic correlations.

Regarding compact-object mergers, involving black holes (BHs) and/or neutron stars (NSs), our analysis suggests that metallicity is either uninfluential or provides a minor contribution to the intrinsic duration vs redshift relation of SGRBs. The main quantity that metallicity could potentially impact, which might affect the intrinsic duration, is the mass of the binary components. However, the masses of the binary components are determined at the time of formation of the binary system and not at the time of their merging (assuming no mass loss/gain), and hence the metallicity of the progenitor stars is what one should keep in mind rather than the metallicity at the time of merging. Indeed, studies show that metallicity could differ between the formation of the binary system and the merging of the system [28]. The fact that the lifetime of a binary system before merging varies appreciably among different systems (even of the same type) illustrates that the connection between intrinsic duration and metallicity via redshift is a complex issue to decipher. Despite this difficulty, many recent studies have investigated the effect of the host galaxy's average metallicity on different merging systems [28]-[34]. A general agreement among these studies is that metallicity does affect the formation of BH-BH and BH-NS binary systems, but not NS-NS binary systems—as thoroughly highlighted and underscored by [34]. Since it is well-known that SGRBs could emanate from NS-NS mergers [35], then this, alongside limitations on GRB detections, makes the effect of metallicity on SGRBs' intrinsic durations less intuitive to understand and establish. Perhaps a way to properly study this dependency is to classify SGRBs based on their type of progenitor (BH-BH, BH-NS, NS-NS or others) and study the $T_{int}$ vs $z$ for each case. This requires multi-messenger detections and proper classifications, areas that researchers are currently working on.

## 4. Conclusion

To sum up, in this study, we investigated the relation between the intrinsic duration and the redshift for both LGRBs and SGRBs. Prior studies have found evidence that the intrinsic duration of long GRBs is inversely correlated with redshift, which is something that our current study confirms using an enlarged dataset. This anticorrelation is most probably due to the influence of metallicity, though the precise mechanism involved needs further investigation. On the other hand, our study finds no evidence for an anticorrelation between the intrinsic duration and redshift for short bursts, which brings into question the effect of metallicity and the extent of its influence on the intrinsic properties of short GRBs, though this is a problem worthy of deeper future study. Moreover, regarding long bursts, future studies might carefully examine and further probe how metallicity affects the physics involved in core-collapse progenitor models and thus enable us to better understand how this anticorrelation for LGRBs arises physically.

## Conflicts of Interest

The authors declare no conflicts of interest regarding the publication of this paper.

## Appendix

**Table A1.** The data for the long GRBs that we compiled from different sources: 1) Reference [20], 2) http://swift.gsfc.nasa.gov/archive/grb_table/,
3) http://user-web.icecube.wisc.edu/~grbweb_public/, 4) Reference [24], 5) Reference [25], 6) Reference [26]. Note that five GRBs had published values of their intrinsic durations, however, the sources did not provide the redshift values for these bursts, so although we included these bursts in our compiled table below for the sake of completeness, we obviously excluded them from our analysis.

| GRB | $z$ | $T_{int}$ (s) | Source |
| --- | --- | --- | --- |
| 180325A | 2.248 | 3.08 | 1, 2 |
| 171222A | 2.409 | 23.58 | 1, 2 |
| 171010A | 0.3293 | 80.74 | 1, 3 |
| 170705A | 2.01 | 7.57 | 1, 2 |
| 170607A | 0.557 | 14.51 | 1, 2 |
| 170214A | 2.53 | 34.81 | 1, 3 |
| 170113A | 1.968 | 16.56 | 1, 2 |
| 161129A | 0.645 | 21.94 | 1, 2 |
| 161117A | 1.549 | 47.93 | 1, 2 |
| 161023A | 2.708 | 13.48 | 1, 3 |
| 161017A | 2.0127 | 10.71 | 1, 2 |
| 161014A | 2.823 | 9.58 | 1, 2 |
| 160804A | 0.736 | 75.8 | 1, 2 |
| 160629A | 3.332 | 14.95 | 1, 3 |
| 160625B | 1.406 | 188.97 | 1, 3 |
| 160623A | 0.367 | 15.77 | 1, 3 |
| 160509A | 1.17 | 170.36 | 1, 3 |
| 160131A | 0.972 | 55.77 | 1, 2 |
| 151027A | 0.81 | 68.17 | 1, 2 |
| 151021A | 2.33 | 17.12 | 1, 2 |
| 150821A | 0.755 | 58.93 | 1, 2 |
| 150727A | 0.313 | 37.63 | 1, 2 |
| 150514A | 0.807 | 5.98 | 1, 3 |
| 150403A | 2.06 | 7.28 | 1, 2 |
| 150323A | 0.593 | 96.33 | 1, 2 |
| 150314A | 1.753 | 3.88 | 1, 2 |
| 150301B | 1.5169 | 5.29 | 1, 2 |
| 150206A | 2.087 | 11.38 | 1, 2 |
| 141225A | 0.915 | 29.41 | 1, 2 |
| 141221A | 1.452 | 9.71 | 1, 2 |
| 141220A | 1.3195 | 3.28 | 1, 2 |

Continued

| | | | |
|---|---|---|---|
| 141028A | 2.33 | 9.46 | 1, 3 |
| 141004A | 0.573 | 1.63 | 1, 2 |
| 140907A | 1.21 | 16.22 | 1, 2 |
| 140808A | 3.29 | 1.04 | 1, 3 |
| 140801A | 1.32 | 3.09 | 1, 3 |
| 140703A | 3.14 | 20.28 | 1, 2 |
| 140623A | 1.92 | 38.05 | 1, 3 |
| 140620A | 2.04 | 15.07 | 1, 3 |
| 140606B | 0.384 | 16.46 | 1, 3 |
| 140512A | 0.725 | 85.78 | 1, 2 |
| 140508A | 1.0285 | 21.85 | 1, 3 |
| 140506A | 0.889 | 33.95 | 1, 2 |
| 140423A | 3.26 | 22.36 | 1, 2 |
| 140419A | 3.956 | 10.78 | 1, 2 |
| 140304A | 5.283 | 4.97 | 1, 2 |
| 140213A | 1.2076 | 8.44 | 1, 2 |
| 140206A | 2.73 | 39.33 | 1, 2 |
| 131231A | 0.642 | 19.02 | 1, 3 |
| 131108A | 2.4 | 5.35 | 1, 3 |
| 131105A | 1.686 | 41.94 | 1, 2 |
| 131011A | 1.874 | 26.81 | 1, 3 |
| 131030A | 1.293 | 6.86 | 1, 2 |
| 130925A | 0.347 | 160.03 | 1, 2 |
| 130907A | 1.238 | 80.49 | 1, 2 |
| 130831A | 0.4791 | 11.85 | 1, 2 |
| 130702A | 0.145 | 51.42 | 1, 3 |
| 130701A | 1.155 | 1.69 | 1, 2 |
| 130612A | 2.006 | 2.47 | 1, 2 |
| 130610A | 2.092 | 7.04 | 1, 2 |
| 130518A | 2.488 | 13.93 | 1, 2 |
| 130505A | 2.27 | 4.49 | 1, 2 |
| 130427A | 0.3399 | 103.17 | 1, 2 |
| 130420A | 1.297 | 45.7 | 1, 2 |
| 130408A | 3.758 | 0.89 | 1, 2 |
| 130215A | 0.597 | 90.01 | 1, 2 |
| 121211A | 1.023 | 2.78 | 1, 2 |
| 121128A | 2.2 | 5.42 | 1, 2 |

Continued

| | | | |
|---|---|---|---|
| 120909A | 3.93 | 22.73 | 1, 2 |
| 120907A | 0.97 | 2.92 | 1, 2 |
| 120811C | 2.671 | 3.91 | 1, 2 |
| 120716A | 2.486 | 68 | 1, 3 |
| 120711A | 1.8037 | 18.31 | 1, 3 |
| 120624B | 1.3487 | 84.87 | 1, 3 |
| 120119A | 1.728 | 20.27 | 1, 2 |
| 111228A | 0.716 | 58.25 | 1, 2 |
| 110918A | 0.982 | 9.87 | 1, 3 |
| 110818A | 3.36 | 15.38 | 1, 2 |
| 110731A | 2.83 | 1.95 | 1, 2 |
| 110715A | 0.82 | 1.56 | 1, 2 |
| 110503A | 1.613 | 2.55 | 1, 2 |
| 110422A | 1.77 | 8.05 | 1, 2 |
| 110213A | 1.46 | 13.95 | 1, 2 |
| 110128A | 2.339 | 3.64 | 1, 2 |
| 110106B | 0.618 | 21.95 | 1, 3 |
| 101219B | 0.718 | 32.91 | 1, 2 |
| 101213A | 0.414 | 21.66 | 1, 3 |
| 100906A | 1.727 | 40.56 | 1, 2 |
| 100814A | 1.44 | 61.69 | 1, 2 |
| 100728B | 2.106 | 3.3 | 1, 2 |
| 100728A | 1.567 | 64.43 | 1, 2 |
| 100621A | 0.542 | 30.04 | 1, 2 |
| 100615A | 1.398 | 15.59 | 1, 2 |
| 100606A | 1.5545 | 23.16 | 1, 4 |
| 100414A | 1.368 | 11.19 | 1, 3 |
| 091208B | 1.0633 | 6.05 | 1, 2 |
| 091127 | 0.49034 | 5.84 | 1, 2 |
| 091024 | 1.091 | 44.91 | 1, 2 |
| 091020 | 1.71 | 8.95 | 1, 2 |
| 091003A | 0.8969 | 10.66 | 1, 3 |
| 090926B | 1.24 | 24.8 | 1, 2 |
| 090926A | 2.1062 | 4.43 | 1, 3 |
| 090902B | 1.822 | 6.85 | 1, 3 |
| 090812 | 2.452 | 9.5 | 1, 2 |
| 090715B | 3 | 17.83 | 1, 2 |
| Continued | | | |
| 090709A | 8.5 | 27.62 | 1, 3 |
| 090618 | 0.54 | 72.98 | 1, 2 |
| 090516A | 4.1045 | 24.09 | 1, 3 |
| 090424 | 0.544 | 9.16 | 1, 2 |
| 090423 | 8 | 0.77 | 1, 2 |
| 090328 | 0.736 | 35.54 | 1, 3 |
| 090323 | 3.57 | 29.58 | 1, 3 |
| 090201 | 2.1 | 21.71 | 1, 3 |
| 090102 | 1.547 | 5.99 | 1, 2 |
| 081222 | 2.77 | 5.01 | 1, 2 |
| 081221 | 2.26 | 9.11 | 1, 2 |
| 081121 | 2.512 | 11.96 | 1, 2 |
| 081109 | 0.9787 | 29.5 | 1, 3 |
| 080916A | 0.689 | 27.44 | 1, 2 |
| 080916C | - | 11.45 | 1 |
| 080721 | 2.591 | 5.5 | 1, 2 |
| 080607 | 3.036 | 7.11 | 1, 2 |
| 080605 | 1.6398 | 5.21 | 1, 2 |
| 080603B | 2.69 | 3.4 | 1, 2 |
| 080514B | - | 2.04 | 1 |
| 080413A | 2.433 | 5.2 | 1, 2 |
| 080411 | 1.03 | 21.11 | 1, 5 |
| 080319C | 1.95 | 3.47 | 1, 2 |
| 080319B | 0.937 | 22.97 | 1, 2 |
| 071117 | 1.331 | 0.98 | 1, 2 |
| 071112C | 0.823 | 2.88 | 1, 2 |
| 071020 | 2.142 | 0.87 | 1, 2 |
| 071010B | 0.947 | 4.37 | 1, 2 |
| 071003 | 1.1 | 8.22 | 1, 2 |
| 070521 | 0.553 | 11.77 | 1, 2 |
| 070508 | 0.82 | 8.1 | 1, 2 |
| 070328 | 0.87 | 17.58 | 1, 5 |
| 070125 | - | 48.71 | 1 |
| 061222A | 2.088 | 19.47 | 1, 2 |
| 061121 | 1.314 | 7.69 | 1, 2 |
| 061021 | 0.3463 | 13.83 | 1, 2 |
| 061007 | 1.261 | 25.47 | 1, 2 |

| Continued | | | |
|---|---|---|---|
| 060814 | 0.84 | 49.94 | 1, 2 |
| 060614 | 0.125 | 109.42 | 1, 2 |
| 060502A | 1.51 | 4.23 | 1, 2 |
| 060124 | 2.297 | 72.16 | 1, 2 |
| 051109A | 2.346 | 11.83 | 1, 2 |
| 051022 | 0.807 | 84.97 | 1, 6 |
| 051008 | 0.94 | 55.4 | 1, 5 |
| 050922C | 2.198 | 1.95 | 1, 2 |
| 050820A | 2.6147 | 122.11 | 1, 2 |
| 050603 | 2.82 | 2.93 | 1, 5 |
| 050525A | 0.606 | 4.38 | 1, 2 |
| 050401 | 2.9 | 8.48 | 1, 2 |
| 041006 | 0.712 | 3.98 | 1, 6 |
| 030329 | 0.168 | 18.64 | 1, 6 |
| 020819B | 0.411 | 5.81 | 1, 6 |
| 020813 | 1.255 | 39.74 | 1, 6 |
| 020405 | 0.69 | 24.58 | 1, 3 |
| 011121 | 0.36 | 41.92 | 1, 3 |
| 010921 | 0.45 | 13.65 | 1, 3 |
| 010222 | 1.477 | 36.2 | 1, 3 |
| 000926 | 2.066 | 17.99 | 1, 3 |
| 000911 | 1.058 | 11.34 | 1, 6 |
| 000418 | 1.118 | 13.09 | 1, 3 |
| 000301C | 2.03 | 1.21 | 1, 3 |
| 000210 | 0.8463 | 4.31 | 1, 3 |
| 000131 | - | 17.54 | 1 |
| 991216 | - | 7.19 | 1 |
| 991208 | 0.706 | 36.88 | 1, 6 |
| 990712 | 0.434 | 11.63 | 1, 3 |
| 990705 | 0.842 | 18.06 | 1, 3 |
| 990510 | 1.619 | 21.33 | 1, 3 |
| 990506 | 1.3 | 55.67 | 1, 3 |
| 990123 | 1.6 | 23.85 | 1, 3 |
| 971214 | 3.42 | 3.6 | 1, 3 |
| 970828 | 0.96 | 33.78 | 1, 6 |
| 970228 | 0.695 | 31.62 | 1, 3 |

Table A2. The data sample of short GRBs that we used in our analysis, taken from [27]. Note that a few of these bursts have an intrinsic duration that exceeds 2 s, but the authors of [27] justified their inclusion as short GRBs because they exhibit features associated with merger systems.

| GRB | $z$ | $T_{int}$ (s) |
|---|---|---|
| 211227A | 0.23 | 3.26 |
| 211211A | 0.08 | 12.08 |
| 210726A | 0.22 | 0.32 |
| 210323A | 0.73 | 0.58 |
| 201221D | 1.05 | 0.07 |
| 200826A | 0.75 | 0.65 |
| 200522A | 0.55 | 0.4 |
| 200411A | 1.93 | 0.48 |
| 200219A | 0.48 | 0.74 |
| 191031D | 1.93 | 0.09 |
| 180805B | 0.66 | 0.6 |
| 180727A | 1.95 | 0.37 |
| 180618A | 0.52 | 2.44 |
| 180418A | 1.56 | 1 |
| 170817A | 0.01 | 2.03 |
| 170728B | 1.49 | 28.6 |
| 170728A | 0.62 | 0.5 |
| 170428A | 0.45 | 0.1 |
| 170127B | 2.28 | 0.53 |
| 161001A | 0.67 | 1.34 |
| 160821B | 0.16 | 0.94 |
| 160624A | 0.48 | 0.26 |
| 160411A | 0.82 | 0.37 |
| 160410A | 1.72 | 0.58 |
| 160408A | 1.90 | 0.36 |
| 151229A | 0.63 | 2.12 |
| 150831A | 1.09 | 0.48 |
| 150728A | 0.46 | 0.57 |
| 150424A | 0.30 | 0.21 |
| 150423A | 1.39 | 0.09 |
| 150120A | 0.46 | 2.28 |
| 150101B | 0.13 | 0.07 |
| 141212A | 0.60 | 0.18 |
| 140930B | 1.47 | 0.41 |

| Continued | | |
|---|---|---|
| 140903A | 0.35 | 0.22 |
| 140622A | 0.96 | 0.07 |
| 140619B | 2.67 | 0.77 |
| 131004A | 0.72 | 0.67 |
| 130822A | 0.15 | 0.03 |
| 130716A | 2.20 | 0.24 |
| 130603B | 0.36 | 0.05 |
| 130515A | 0.80 | 0.14 |
| 120804A | 1.30 | 0.35 |
| 120305A | 0.23 | 0.08 |
| 111117A | 2.21 | 0.13 |
| 101224A | 0.45 | 1.19 |
| 101219A | 0.72 | 0.3 |
| 100816A | 0.80 | 1.13 |
| 100625A | 0.45 | 0.17 |
| 100206A | 0.41 | 0.13 |
| 100117A | 0.92 | 0.13 |
| 091117A | 0.10 | 0.24 |
| 090227B | 1.61 | 0.12 |
| 081024A | 3.05 | 0.16 |
| 070714B | 0.92 | 0.65 |
| 070429B | 0.90 | 0.26 |
| 060502B | 0.29 | 0.07 |
| 051221A | 0.55 | 0.14 |
| 050509B | 0.23 | 0.07 |
| 090927 | 1.37 | 0.22 |
| 090515 | 0.40 | 0.03 |
| 090510 | 0.90 | 0.5 |
| 090426 | 2.61 | 0.33 |
| 080905 | 0.12 | 0.86 |
| 080123 | 0.50 | 76.92 |
| 071227 | 0.38 | 0.52 |
| 070809 | 0.22 | 1.07 |
| 070729 | 0.52 | 0.59 |
| 070724 | 0.46 | 0.27 |
| 061217 | 0.83 | 0.16 |
| 061210 | 0.41 | 0.07 |

Continued

| | | |
|---|---|---|
| 061201 | 0.11 | 0.52 |
| 061006 | 0.44 | 0.26 |
| 060801 | 1.13 | 0.23 |
| 060614 | 0.13 | 5.33 |
| 060121 | 4.60 | 0.36 |
| 051227 | 0.80 | 2.39 |
| 051210 | 2.58 | 0.39 |
| 050813 | 0.72 | 0.35 |
| 050724 | 0.26 | 1.99 |
| 050709 | 0.16 | 0.06 |
| 050709 | 0.1606 | 0.06 |
| 050509B | 0.226 | 0.07 |